# Robust Fourier ptychographic microscopy via a physics-based defocusing strategy for calibrating angle-varied LED illumination

CHUANJIAN ZHENG,[1] SHAOHUI ZHANG,[1,2] GUOCHENG ZHOU,[1] YAO HU[1], AND QUN HAO[3]

[1]*School of Optics and Photonics, Beijing Institute of technology, Beijing 100081, China*
[2]*zhangshaohui@bit.edu.cn*
[3]*qhao@bit.edu.cn*

**Abstract:** Fourier ptychographic microscopy (FPM) is a recently developed computational imaging technique for wide-field, high-resolution microscopy with a high space-bandwidth product. It integrates the concepts of synthetic aperture and phase retrieval to surpass the resolution limit imposed by the employed objective lens. In the FPM framework, the position of each sub-spectrum needs to be accurately known to ensure the success of the phase retrieval process. Different from the conventional methods with mechanical adjustment or data-driven optimization strategies, here we report a physics-based defocusing strategy for correcting large-scale positional deviation of the LED illumination in FPM. Based on a subpixel image registration process with a defocused object, we can directly infer the illumination parameters including the lateral offsets of the light source, the in-plane rotation angle of the LED array, and the distance between the sample and the LED board. The feasibility and effectiveness of our method are validated with both simulation study and experiments. We show that the reported strategy can obtain high-quality reconstruction of both the complex object and pupil even the LED array is randomly placed under the sample with both unknown lateral offsets and rotations. As such, it enables the development of robust FPM systems by reducing the requirement on fine mechanical adjustment and data-driven correction in the construction process.

## 1. Introduction

Fourier ptychography (FP) [1-3] is a recently developed computational imaging technique, which addresses the inherent trade-off between large field-of-view (FOV) and high spatial resolution in conventional optical systems. Based on the corresponding relationship between the illumination direction and the spatial spectrum position, FP can obtain an expanded range of spectrum by providing varying illumination angles, and achieve quantitative phase imaging with a gigapixels space-bandwidth product (SBP) [4] by integrating phase retrieval [5] and synthetic aperture [6] algorithms.

Since its first demonstration in 2013, FP has evolved from a microscopic imaging tool to a versatile imaging technique, embodied in different implementations including reflective FP [7, 8], aperture-scanning FP [9], X-ray FP [10], multi-camera FP [11], among others. Meanwhile, FP has been applied in various applications, including quantitative 3D imaging [12-14], digital pathology [15, 16], aberration metrology [17] and extension for incoherent imaging [18], et. al.

In a typical FPM system, an LED array is used to provide angle-varied illumination for the sample. During the data acquisition process, LEDs are turned on sequentially to provide different angular illuminations and a camera is used to record the corresponding low resolution (LR) images whose spatial frequency ranges are determined by the NA of the objective lens and the illumination wavelength and wavevectors. In the original FPM model, each LED element is assumed to be a point source that emits quasi-monochromatic light, and the whole LED board should be aligned precisely. The former one can be achieved by choosing LED element with a narrower spectral bandwidth and a smaller light-emitting area. Nevertheless, the positional deviation of the LED board is unavoidable during the process of constructing or modifying FPM systems, and it will degrade the quality of the reconstruction by making the sub-spectrum position used in the phase retrieval algorithm inconsistent with the wavevector corresponding to the actual illumination directions. One conventional way to deal with this problem is to pre-calibrate the LED array by mechanical adjustment stages [19], but aligning the LED array accurately is a time-consuming task and needs precise and multi-degree of freedom stages, which are always bulky and expensive. An intuitive and effective method [20] utilizing brightfield-to-darkfield features to get the location and orientation of the LED array is adjust-free, and no additional hardware or operations are required. Nevertheless, it may fail when there are no expected brightfield overlapping zone for some specific system parameters, such as the adjacent LED distance, the objective NA, and the distance between the sample and the LED array. Many data-driven optimization approaches taking use of the LR data redundancy are also useful solutions, such as pcFPM method [21], two-part calibration method [22], QG-FPM [23], and tcFPM [24], et.al. However, all these methods use the intensity distribution characteristics of the LR images to seek an optimal positional parameters estimation, whereas besides the positional deviation, the LR images' intensity distributions are also affected by illumination intensity fluctuation, objective aberrations, random noise and other kind of systematic errors. The ways in which all errors affect the LR images are mixed together, result in extracting positional deviation from the LR images constraint become an ill-posed problem. It is thus difficult for these methods to be performed with the optimization and correction methods of other non-ideal parameters. In addition, the data-driven optimization approaches also have the problem of time-consuming and easy to fall into local traps.

To eliminate the influence of other non-ideal systematic parameters and accurately obtain the positional deviation of the LED board, we propose a correction method with an explicit imaging physical model, called pdcFPM. Based on the constraint that each LED element is arranged in a regular array, and the relationship between the lateral offset of the defocus sample image and the illumination direction [25-27], the positional parameters of the LED board can be calculated precisely with the aid of subpixel image registration and non-linear regression algorithms. Simulations and experiments have been conducted to demonstrated the feasibility and effectiveness of pdcFPM for large scale positional deviations. The proposed method can significantly improve the quality of the reconstructed complex amplitude images, obtain the objective pupil function without influence of positional deviations, and therefore evidently reduce the accuracy requirements of the LED array board in the process of building FPM platforms.

The remainder of this paper is organized in the following manner. In Section 2.1, we analyze the effect of LED positional deviation through the reconstructed results, in Section 2.2, we establish a mathematical model between LED positional parameters and LR image offset, and in Section 2.3, we give the flow process of the method. The effectiveness of pdcFPM is confirmed with simulations and experiments in Sections 3 and 4, respectively, and the summary and further discussion are given in Section 5.

## 2. Principle

### 2.1 Positional deviation in FPM system

A typical FPM system generally consists of an LED array, a low-NA objective lens, and a monochromatic camera. When acquiring the raw data set, the LEDs are turned on in a sequential manner to provide angular-varying illuminations. When $LED_{m,n}$ ($m$, n are the indexes of rows and columns) is working, an plane light with a wave vector of $(u_{m,n}, v_{m,n})$ is generated to illuminate for a small segment of the sample, and the camera is used to acquire the corresponding LR intensity image $I_{m,n}^c(x',y')$. Subsequently, these raw images are processed by the conventional FP reconstruction algorithm to obtain the HR complex amplitude information of the sample. The reconstruction process is divided into five steps. The first step is to initialize of the estimation of sample complex amplitude $o_0(x,y)$ and the pupil function $P_0(u,v)$, then the spectrum estimation of sample is $O_0(u,v)=\mathcal{F}\{o_0(x,y)\}$, where '$\mathcal{F}$' denotes the Fourier transform. Secondly, using the current estimation of pupil function and spectrum of sample to obtain the LR image estimation $o_{0,m,n}^e(x,y)=\mathcal{F}^{-1}(O_0(u-u_{m,n},v-v_{m,n})P_0(u,v)$, where '$\mathcal{F}^{-1}$' represents the inverse Fourier transform. Thirdly, replacing the amplitude of $o_{0,m,n}^e(x,y)$ with the captured LR image amplitude $\sqrt{I_{m,n}^c(x',y')}$ and keep the phase unchanged, then the updated LR image estimate $o_{0,m,n}^u(x,y)$ can be obtained. The fourth step is to update the corresponding sub-spectrum of the HR spectrum using the spectrum of $o_{0,m,n}^u(x,y)$. The last step is to repeat the above four steps until all the LR images are used. The whole process will be repeated $J$ times until the solution converges and the HR complex amplitude image $o_{J,m,n}^e(x,y)$ is formed.

It is worth noting that the accurate position of each sub-spectrum needs to be known when updating the HR spectrum in the fourth step for high recovery quality. In other words, the illumination wave vectors determined by the position of each LED element and the distance between LED board and sample need to be precisely known, which put high requirements on reconstructing and modifying FPM systems. Fortunately, some improved recovery strategies have been proposed to relax the demands. EPRY algorithm [28] is a helpful method to get better solution by updating HR spectrum and pupil function simultaneously. When the deviation between ideal and actual positions of sub-spectrum is not very large, it can be corrected by EPRY algorithm. However, as the deviation becomes greater, the performance of EPRY algorithm will decrease, and resulting in wrinkles artifacts in the reconstructed amplitude and phase, the quality of the result is seriously degraded.

Here, we demonstrate the influences of positional deviation of LED array by simulation. For directly analyzing the relationship between recovery quality and positional deviation, we assume the LED array is only moved along the *x*-axis, which is a pre-defined direction in the transverse plane, and use $\Delta x$ to represent the degree of deviation. We use EPRY algorithm to recovery the intensity and phase when $\Delta x$ varies from 0 to 2000$\mu m$, and the results is shown in Figure 1. Figure. 1(a1) and (b1) shows the intensity and phase ground truths of the sample. When $\Delta x$ is equal to 200$\mu m$, the reconstructed intensity and phase are shown in Figure. 1(a2) and (b2) respectively. Although the images are blurred and the center of the phase becomes black, the change is not disastrous, and distribution characteristic of both the intensity and phase are still distinguishable. When $\Delta x$ increases to 800$\mu m$, the reconstructed intensity and phase are shown in Fig. 1(a3) and (b3), respectively. Although most of the intensity information can still be distinguished, wrinkle artifacts appear in the intensity and phase images and the phase information has been seriously distorted. When $\Delta x$ reaches 2000$\mu m$ which has resulted in the wrong choice of center LED, obviously black artifacts and some unexpected phase information arise in intensity image as shown in Fig. 1(a4), and the phase information shown in Fig. 1(b4) is more terrible because whole image is completely submerged and indistinguishable.

By intuitively observing the HR reconstructed results with the same $\Delta x$, it can be found that the reconstructed phase quality degradation is more serious compared to the reconstructed intensity. That is because the sub-spectrum error in the Fourier domain is directly related to the phase information in the spatial domain. Figure.1 (c1) and (c2) gives the relationship between the root mean square error (RMSE) of the reconstructed and real intensity and phase and the positional deviation $\Delta x$. The RMSE of the phase is obviously larger than intensity. In addition, it is noteworthy that the RMSE is not linearly related to $\Delta x$, but remains stable when $\Delta x$ changes in a certain interval (e.g., the RMSE of intensity is about 0.2 when $\Delta x$ varies from 1400$\mu m$ to 1800$\mu m$), and the RMSE increases sharply when $\Delta x$ exceeds a certain threshold (e.g., 800$\mu m$). The abnormal phenomenon is caused by the discretization of the Fourier domain coordinates when reconstructing the HR image. If the deviation between the real and ideal positions of sub-spectrum is no more than one pixel in Fourier domain, the reconstruction quality of the images is stable and the RMSE will not change significantly. However, if the deviation is more than one pixel, then the recovery quality will start to drop sharply.

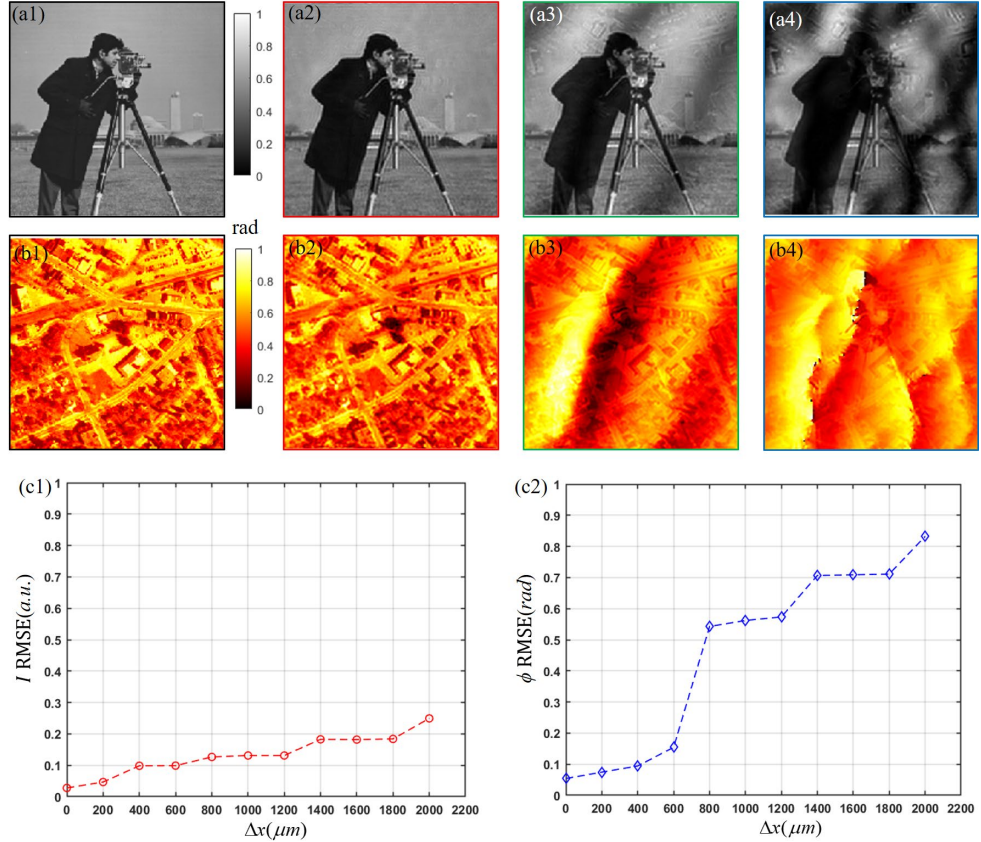

Fig. 1. Quality of reconstructed results versus LED array positional deviation. (a1) and (b1) are the amplitude and phase profiles of the sample; (a2)-a(4) are the reconstructed image intensities when $\Delta x$ is equal to 200$\mu m$, 800$\mu m$ and 2000$\mu m$, respectively; (b2)-b(4) are the corresponding reconstructed phase images; (c1) and (c2) are the RMSE of reconstructed intensity and phase versus $\Delta x$.

Moreover, in an actual FPM system, the LED array will also have other positional deviations besides the *x*-axis shift, such as the y-axis shift and the rotation along optical axis, et.al. The recovery quality will degrade more seriously under the combined effect of these deviations, so it is very significant to correct the positional deviation of the LED array.

*2.2 Image Offset Model*

A typical FPM platform and the corresponding LED array with positional deviation are shown in Fig. 2. The blue LED element in Fig. 2(c) is considered as the center LED but point *O* is the actual center with respect to the sample, and the orientation of LED array (the black dashed line) deviates from the *x*-axis. In fact, the high precision of LED board manufactured by existing technique enables us to consider the arrangement of LED elements is regular, and FPM system is always placed on a horizontal tale. We can assume the LED array shown in Figure 2(b) is perpendicular to the optical axis $z$ to simplify the LED based illumination model. Then, the positional coordinates of each LED element can be express as

$$\begin{aligned} x_{m,n} &= [m\cos(\theta) + n\sin(\theta)]d_{LED} + \Delta x, \\ y_{m,n} &= [-n\sin(\theta) + n\cos(\theta)]d_{LED} + \Delta y, \end{aligned} \quad (1)$$

where ($x_{m,n}$, $y_{m,n}$) represents the Cartesian coordinate of the LED element on the row $m$, column $n$, $\theta$ denotes the rotation angle of the LED array around the optical axis $z$, $d_{LED}$ is the distance between adjacent LED elements, $\Delta x$ and $\Delta y$ represent the shift of the LED array along the x-axis and y-axis, respectively.

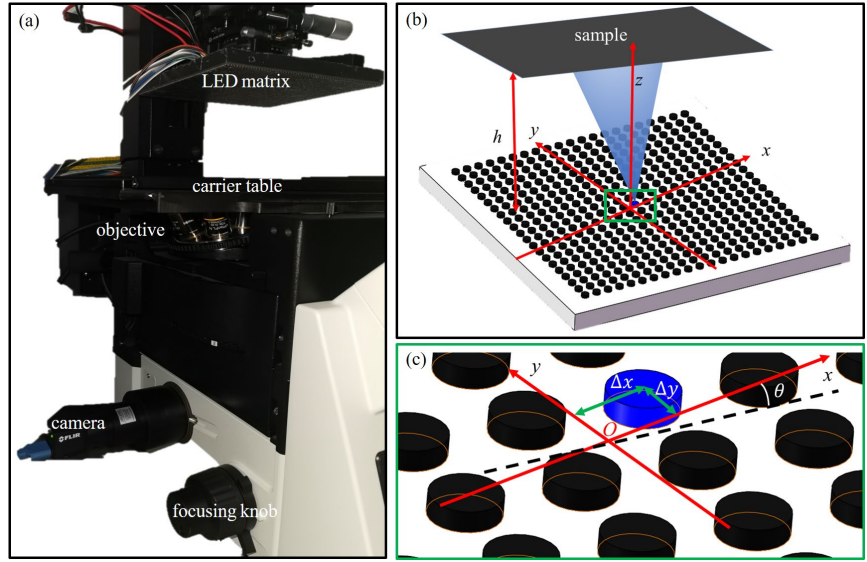

Fig. 2. FPM system and LED array model. (a) is an FPM system; (b) is the corresponding LED array model with positional deviation; (c) is the local magnification of (b).

When turning on $LED_{m,n}$ to provide illumination, the illumination light can be considered as a plane wave for each small segment of the sample, and the wave-vector $(u_{m,n}, v_{m,n})$ is

$$\begin{aligned} u_{m,n} &= -\frac{2\pi}{\lambda} \cdot \frac{x_c - x_{m,n}}{\sqrt{(x_c - x_{m,n})^2 + (y_c - y_{m,n})^2 + h^2}}, \\ v_{m,n} &= -\frac{2\pi}{\lambda} \cdot \frac{y_c - y_{m,n}}{\sqrt{(x_c - x_{m,n})^2 + (y_c - y_{m,n})^2 + h^2}}, \end{aligned} \quad (2)$$

where $\lambda$ denotes the wavelength of the illumination, $(x_c, y_c)$ is the center coordinate of each small segment, and $h$ is the distance between the sample and the LED array.

When the sample is in the focus position, the LR image acquired by the camera can be expressed as

$$I_{m,n}^c(x',y') = \mathcal{F}^{-1}\{O(u - u_{m,n}, v - v_{m,n}) \cdot P(u,v)\}, \quad (3)$$

where $(x', y')$ is the Cartesian coordinate of the camera plane, and $P(u,v)$ is the coherent transfer function of the FPM system.

When the sample is in the defocus position, the LR image can be described as

$$I^c_{m,n,d}(x',y') = \mathcal{F}^{-1}\{O(u-u_{m,n}, v-v_{m,n}) \cdot P(u,v)H(u,v,z_d)\}, \tag{4}$$

where $z_d$ is the defocus distance, and $P(u,v)H(u,v,z_d)$ is the coherence transfer function of the defocus FPM system. $H(u,v,z_d)$ can be expressed as

$$H(u,v,z_d) = \exp[jAz_d(u\frac{u_{m,n}}{w_{m,n}} + v\frac{v_{m,n}}{w_{m,n}})]\exp[jAz_d w_{m,n}]\exp(-jAz_d\frac{u^2+u^2}{2w_{m,n}})], \tag{5}$$

where $A$ is the magnification of the objective lens, and $w_{m,n}$ denotes the wave vector along the z direction and can be calculate as

$$w_{m,n} = \sqrt{(\frac{2\pi}{\lambda})^2 - u^2_{m,n} - v^2_{m,n}}. \tag{6}$$

The first terms in $H(u,v,z_d)$ are phase change in Fourier domain and will result in offset in space domain, and the third term will make the quality of the LR image degrade. We take no account of the effect caused by the third term and derive the relationship between the defocus and focus images as

$$\begin{aligned} I^c_{m,n,d}(x',y') &= I^c_{m,n}(x'+\Delta x'_{m,n}, y'+\Delta y'_{m,n}), \\ \Delta x'_{m,n} &= Az_d \frac{u_{m,n}}{w_{m,n}}, \\ \Delta y'_{m,n} &= Az_d \frac{v_{m,n}}{w_{m,n}}, \end{aligned} \tag{7}$$

where $\Delta x'_{m,n}$ and $\Delta y'_{m,n}$ are the offset between the defocus and focus LR images along the $x'$ and $y'$ direction in the image plane, respectively. Because both $A$ and $z_d$ are constants in a fixed system, then the offset depends entirely on ($u_{m,n}, v_{m,n}$) for a certain small segment of sample. Therefore, the illumination direction of each LED element can be obtained by calculating the image offset.

## 2.3 Correction strategy

The process of pdcFPM is shown in Figure 3. Firstly, we adjust the sample to the focus position with the focusing knob as shown in Fig. 2(a), then turn on $LED_{0,0}$ and capture the corresponding LR image $I^c_{0,0}(x',y')$. In FPM system, determining whether the sample is in the focus position or not is an easy task; when the sample is in the focus position, the LR image acquired will not shift regardless of the angle of the incident light because $z_d$ is zero in equation (6). Therefore, if the LR image does not shift with different LED illumination, then we can render the sample is in the focus position. In this step, although it is feasible to choose to any LED element in the bright field as the light source, considering that the light emitted from $LED_{0,0}$ has better symmetry, the LR image $I^c_{0,0}(x',y')$ is better to be the reference image.

Secondly, we adjust the sample defocus to 200$\mu m$ away from the focus position, then turn on the LED elements within the bright field of the objective, and capture the corresponding LR images $I^c_{m,n,d}(x',y'), m,n = -2,1,0,1,2$. In this step, the defocus distance is a crucial parameter. The defocus distance can't be too small, otherwise, the calculated image offset will have a low signal-to-noise ratio because the offset of the image is too small. In contrary, the image will be seriously distorted because of the defocus aberration, and the calculated offset is not accurate. In this paper, we take $z_d = 200\mu m$, where the image offset is enough and the image quality is also high. Furthermore, pdcFPM is based on the accurate calculation of relative image offset, and the geometrical features of BF images are much better than DF ones, so we choose the corresponding LED elements to implement our approach.

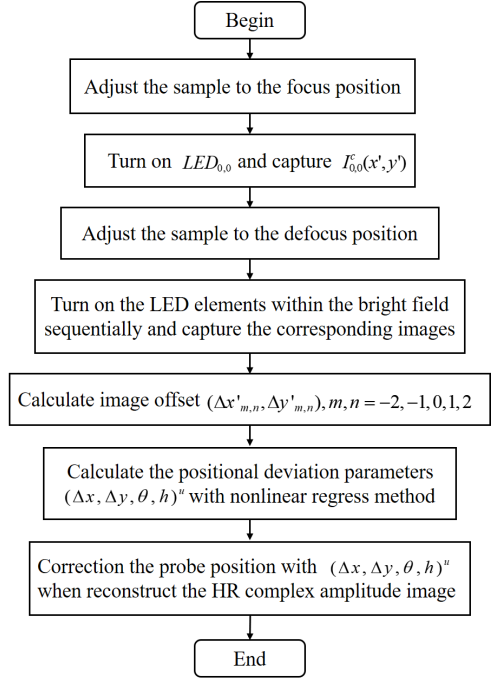

Fig. 3. Block diagram of pdcFPM strategy

Next, we calculate the image offsets $(\Delta x'_{m,n}, \Delta y'_{m,n})$ between the defocus images and the reference image. Considering calculating the offset of the images within the whole FOV is unnecessary and time-consuming, we choose to calculate the offset of a small segment, which can be a stripe on the resolution target, a small hole, and a cell, et al. Furthermore, because the grayscale of background will exert an effect on precision of calculating the image offset, we binarize the selected small segment and then calculate the offset of the centroid with subpixel image registration algorithm.

After getting the image offsets, we utilize non-linear regression algorithm to obtain the positional parameters of the LED array. Mathematically, the non-linear regression process can be described as

$$E(\Delta x, \Delta y, \theta, h) = \sum_{m,n=-2}^{2} \{[\Delta x'_{m,n} - \Delta x'_{m,n,e}(\Delta x, \Delta y, \theta, h)]^2 + [\Delta y'_{m,n} - \Delta y'_{m,n,e}(\Delta x, \Delta y, \theta, h)]^2\}$$
$$(\Delta x, \Delta y, \theta, h)^c = argmin[E(\Delta x, \Delta y, \theta, h)]$$
, (8)

where $E(\Delta x, \Delta y, \theta, h)$ is the defined non-linear regression function which should be minimized, $[\Delta x'_{m,n,e}(\Delta x, \Delta y, \theta, h), \Delta y'_{m,n,e}(\Delta x, \Delta y, \theta, h)]$ denotes the image offset estimation calculated by Eq. (6), and $(\Delta x, \Delta y, \theta, h)^c$ is the optimal estimate of the positional deviation parameters of the LED array.

Finally, we utilize $(\Delta x, \Delta y, \theta, h)^c$ to calculate the corrected sub-spectrum position $(u_{m,n}, v_{m,n})$ in the process of updating HR spectrum. However, when the LED array has a considerable deviation, the original optimal center LED may be replaced by other LEDs around it, and then the convergence of the solution will be degraded if the recovery process is iterated according to the original spiral order. In the process of reconstructing HR image with pdcFPM, we first calculate the distances between the corrected positions $(u_{m,n}, v_{m,n})$ and zero frequency's position, and then rearrange the update order from nearest to furthest according to the distances to ensure fast convergence even under large positional deviations.

## 3. Simulation

Before applying pdcFPM strategy to correct the positional deviation of actual FPM platform, we first validate its effectiveness with simulations. The parameters in the simulations are chosen based on the actual system. We utilize a 21×21 programable LED array as the light source to provide angle-varied illuminations. The distance between adjacent LED elements is 2.5mm, and the wavelength of the illumination light is 470nm. The NA of the objective lens is 0.1, and the magnification is 4. The pixel size of camera is 2.4μm.

As mentioned in section 2.3, we use the offset of a small segment to represent the offset of image. Considering that there are many rectangular stripes on resolution target, we employ a square hole with a side length of 50μm as the sample and place it beneath the LED array and the distance is 92mm, and the defocus distance $z_d$ is 200μm. When simulating the LR images, the positional deviations are introduced artificially and taken in the range of $\Delta x \in (-2.5mm, 2.5mm)$, $\Delta y \in (-2.5mm, 2.5mm)$, $h \in (87mm, 97mm)$, $\theta \in (-5°, 5°)$. In an actual FPM system, the positional deviations can be adjusted through rough adjustment with mechanical adjustment devices. In addition, if the FPM system is not equipped with any mechanical adjustment devices for the LED array, we can choose the optimal LED element as the center one in which the transversal deviation is smaller than the deliberately set maximum value according to the characterstic of the spots acquired by the camera; and even unskilled users choose the wrong LED element as the center one, we can also rearrange the update order when reconstructing the HR images to remain fast convergence and good recovery quality.

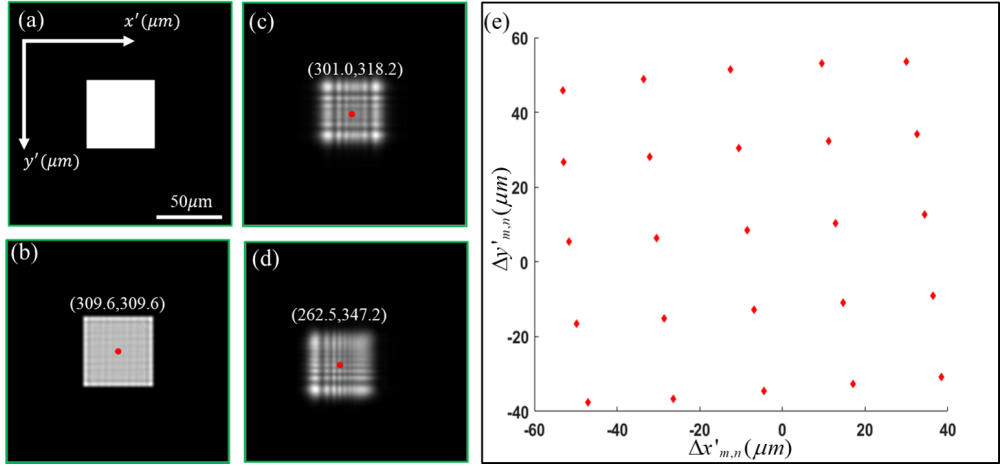

Fig. 4. Simulation results of image offset with the positional parameters $\Delta x=1mm$, $\Delta y=1mm$, $\theta=5°$, and $h=92mm$. (a) is the intensity of sample profile (b) is the LR image $I^c_{0,0}(x',y')$. (c) and (d) are $I^c_{0,0,d}(x',y')$ and $I^c_{2,2,d}(x',y')$ respectively. (e) is the offset between the 25 defocus LR images in the bright field and the focus image.

Figure 4 shows the simulation results of image offset with deviation parameters of $(\Delta x = 1mm, \Delta y = 1mm, \theta = 5°, h = 92mm)$. Figure. 4(a) shows the small square sample that is used to calculate the image offset. Figure. 4(b) is the LR image when the sample is in the focus position and illuminated with the center LED element. Figure. 4(c) shows the image when the sample is in the defocus position and illuminated with the center LED element. Figure. 4(d) shows the image when the sample is in the defocus position and illuminated with the LED element of row 2, column 2. Figure. 4(e) shows the offsets of 25 defocus images relative to the focus image. The parameters obtained by the nonlinear regression is $(\Delta x = 0.997mm, \Delta y = 1.007mm, \theta = 5.243°, h = 92.15mm)$, which can match the actual introduced parameters well.

Table 1. Correction results under large positional deviations

| Actual parameters $\Delta x(mm)$, $\Delta y(mm)$, $\theta(°)$, $h(mm)$ | Corrected parameters $\Delta x^c(mm)$, $\Delta y^c(mm)$, $\theta^c(°)$, $h^c(mm)$ |
|---|---|
| 1.5, 1.5, 5, 92 | 1.494, 1.503, 5.200, 92.20 |
| 2.0, 2.0, 5, 92 | 2.001, 1.999, 4.949, 92.04 |
| 2.5, 2.5, 5, 92 | 2.499, 2.488, 4.951, 92.92 |

To verify the performance of the pdcFPM under other large positional deviations, we continue to perform the simulation with the positional parameters within the setting range. Table 1 shows the simulation results when the LED array has different representative deviation parameters. The parameters obtained by pdcFPM are still in good agreement with the actual parameters.

Next, we utilize the obtained positional parameters to correct the position of the sub-spectrum position in the Fourier domain during the process of reconstructing the HR image. Fig. 5 shows the recovery HR images with conventional FPM and pdcFPM under different positional deviations. Fig. 5(a1) and (a2) is the intensity and phase of the sample. Fig. 5(b1) and (b2) show the reconstructed intensity and phase images with conventional FPM when the introduced parameter is $(\Delta x = 1mm, \Delta y = 1mm, \theta = 5°, h = 92mm)$, obvious black and white artifacts arise in both intensity and phase images. Fig. 5(b3) and (b4) are the HR intensity and phase images with pdcFPM, all disturbed information disappears with the correction of pdcFPM. With the positional parameters increase, the reconstructed intensity and phase imaging quality also continuously degrade, and some regular wrinkles are coupled into the images, as shown in Fig. 5(e1) and (e2). the quality of HR images can still be sufficiently improved with correction. Although under such remarkable positional deviations, pdcFPM can still maintain excellent performance, which illustrates the effectiveness of pdcFPM for large scale positional deviation.

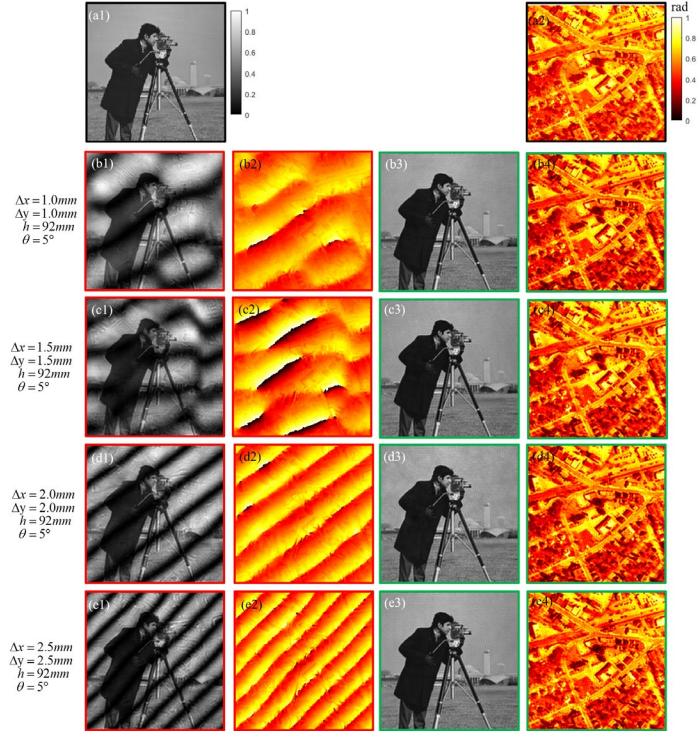

Fig. 5 Simulation results under different positional deviation. Fig. 5(a1) and (a2) are the intensity and phase of the sample. (d1)-(e1) are the HR intensity images with conventional FPM. (d2)-(e2) are the HR phase images with conventional FPM. (b3)-(e3) are the HR intensity images with pdcFPM. (b4)-(d4) are the HR phase images with pdcFPM.

## 4. Experiment

To evaluate the effectiveness of pdcFPM experimentally, we use a USA-1951 resolution target as sample to compare the reconstructed intensity distribution and pupil function with conventional FPM and pdcFPM respectively.

The light source used in our FPM system shown in Fig.2 is a 21×21 LED array, the distance between adjacent LEDs is 2.5*mm*, the illumination wavelength is 470*nm* and its bandwidth is 20*nm*, the NA of the objective lens is 0.1, and the magnification of the objective is 4. We use a camera (FLIR, BFS-U3-200S6M-C, sensor size 1", dynamic range 71.89 dB, pixel size 2.4μm) to record the LR images. Figure 6 shows the performance of EPRY-FPM and pdcFPM with different LED array positional deviations. We use the center-of-mass offset of a small rectangular shown in Fig. 6(b1) to represent the offset of the image for simplifying calculating and accelerating the speed of subpixel image registration algorithm. Fig. 6(b2) is a small segment of Fig. 6(a), the full FOV LR image of the target. Fig. 6(c1) and (c2) show the reconstructed HR intensity image and pupil function without positional deviation. The HR intensity image has very good imaging quality and the reconstructed pupil has good symmetry which is consistent with prior knowledge. Next, we translate the LED array by 2 mm along the y-axis with a mechanical adjustment stage. The corresponding reconstructed intensity image and pupil function with EPRY-FPM are shown in Fig. 6(d1) and (d2), in which wrinkle artifacts appear in the reconstructed intensity image, while the degradation of imaging quality is not severe, which is because EPRY-FPM will identify the positional deviations of the LED array and consider it as a kind of aberration. Therefore, EPRY-FPM couples the positional deviations of the LED array into the reconstructed pupil function, then the pupil in Figure. 6(d2) is distorted and not reliable. In addition, the correction ability of EPRY-FPM is also limited, for example when the positional deviation is remarkably large, the reconstructed results with EPRY-FPM will be significantly worse. Similarly, we use the mechanical adjustment stage to move the LED array along the x and y axis respectively by 2mm, the reconstructed HR intensity image and pupil function with EPRY-FPM shown in Figure 6(f1) and (f2) are severely degraded for the existence of obvious wrinkle artifacts, and the details of the resolution line pair is difficult to distinguish.

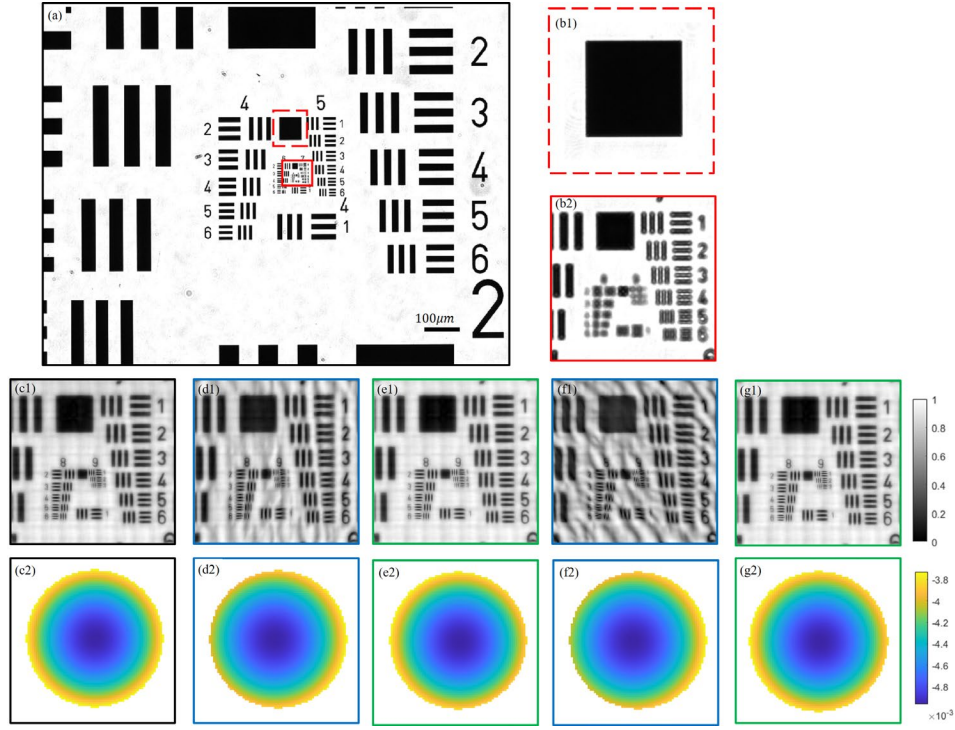

Fig. 6 Performance comparison of EPRY-FPM and pdcFPM. (a) is the full FOV LR image acquired by the camera; (b1) is the small segment used to calculate the image offset; (b2) is the LR image area used for reconstruction; (c1) and (c2) are the reconstructed intensity image and pupil function with EPRY-FPM when there is no positional deviation; (d1) and (d2) are the reconstructed intensity image and pupil function with EPRY-FPM when $\Delta y=2mm$; (e1)-(e2) are the reconstructed intensity image and pupil function with pdcFPM when $\Delta y=2mm$; (f1)-(f2) are the reconstructed intensity image and pupil function with EPRY-FPM when $\Delta x=2mm$ and $\Delta y=2mm$; (g1)-(g2) are the reconstructed intensity image and pupil function with pdcFPM when $\Delta x=2mm$ and $\Delta y=2mm$.

To correct the positional deviations of the LED array, we use pdcFPM to recover the HR images and pupil functions. When the LED array is moved $2mm$ along the $y$-axis, the recovered positional parameters are $\Delta x = -0.279mm$, $\Delta y = 1.980mm$, $\theta = -0.393°$, and $h = 93.39mm$, then we use these parameters to correct the sub-spectrum positions in the process of phase retrieval and HR spectrum synthesis. The reconstructed intensity image and pupil function are shown in Figure.6 (e1) and (e2), respectively. Compared with Fig. 6(f1) and (f2), wrinkle artifacts no longer appear in the reconstructed intensity image and the distortion of pupil function also disappears. When the LED array is shifted by $2mm$ in both the $x$ and $y$ directions, the recovered positional parameters are $\Delta x = 1.949mm$, $\Delta y = 1.997mm$, $\theta = -0.327°$, and $h = 93.04mm$,. As shown in Fig. 6(g1) and (g2), the reconstructed intensity image and pupil function still remain high-quality even with such large positional deviations, demonstrating the effectiveness of pdcFPM.

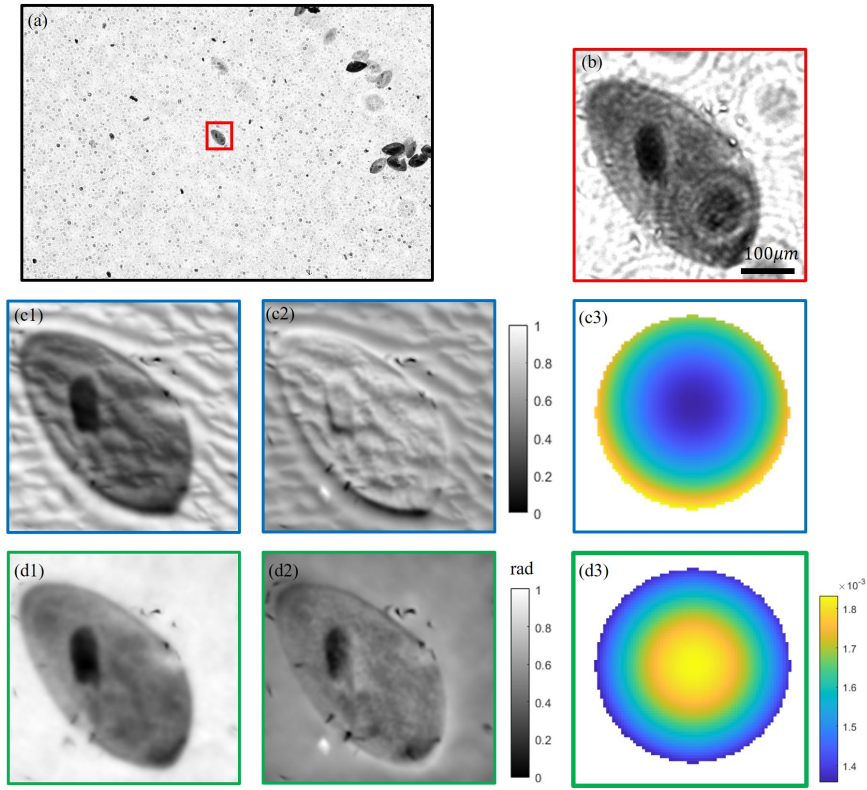

Fig. 7 pdcFPM correction for the random positional parameters. (a) is the full FOV LR image of the paramecium slice; (b) is a small segment of (a); (c1)-(c3) are the reconstructed intensity, phase and pupil with EPRY-FPM; (d1)-(d3) are the reconstructed intensity, phase and pupil with pdcFPM.

In addition, to verify the generalizability of our method, we use paramecium slice as the sample, and randomly move and rotate the LED array to an arbitrary position. Then, we compare the reconstructed intensity, phase, and pupil function results before and after utilizing pdcFPM. Fig. 7(a) shows the full FOV LR image, and Fig. 7(b) is the enlargement of the small segment in the red box in Fig.7(a). The reconstructed intensity, phase, and pupil function with EPRY-FPM are shown in Fig. 7(c1-c3), respectively. Many wrinkle artifacts appeared in both the reconstructed intensity and phase distribution, decrease the reconstructed quality of the image severely; the reconstructed phase image's contrast is severely reduced. In addition, the reconstructed pupil is also severely distorted. Subsequently, the positional parameters of the LED array are calculated with pdcFPM, which are $\Delta x = 1.680 mm$, $\Delta y = 0.089 mm$, $\theta = -5.135°$, and $h = 93.15 mm$, respectively. The reconstructed intensity, phase, and pupil function obtained with these parameters are shown in Fig. 7(d1-d3), respectively. Compared with Fig. 7(c1-c3), the artifacts caused by the positional deviations vanish in both the reconstructed intensity and phase images, and the imaging quality of the image is highly improved.

## 5. Discussion

In this paper, we propose a positional correction method termed pdcFPM that yields high-quality reconstructed intensity, phase, and pupil function. The feasibility and effectiveness of pdcFPM are verified by both simulations and experiments. Different from the existing complex data-driven optimization strategies, pdcFPM constructs a clear physical model that can separate the positional deviations from many other coupled systematic errors effectively. We utilize the relationship between the offset of the defocus image and illumination direction to firstly obtain four key positional parameters of each LED, and then precisely calibrate the position of sub-spectrum in the process of phase retrieval and aperture synthesis. The comparative experimental results corresponding to large scale positional deviations prove the excellent performance of pdcFPM. In addition, even if the positional deviations are remarkably large (larger than the interval of adjacent LEDs) that can cause wrong choice of the center LED, our method can still achieve a good recovery results by rearranging the updating order. Thus, our method can handle arbitrary positional deviations.

Although pdcFPM can correct large scale positional parameters and reduce the complexity of system construction, the accuracy of the focusing knob used for defocus adjustment is highly demanded because the image offset is as small as tens of microns and easily affected. For the focusing device with poor accuracy, it can also be feasible because the image offset caused by the mechanical adjustment can be calibrated before the experiment. To further reduce the requirements on the hardware system and eliminate the defocus adjustment part, we are trying to implement LED array positional correction based on the bright-dark field transition boundary characteristics.

**Funding.** National Natural Science Foundation of China (61735003, 61805011).

**Disclosures.** The authors declare no conflicts of interest.

**Data availability.** Data underlying the results presented in this paper are not publicly available at this time but may be obtained from the authors upon reasonable request.

**Acknowledgements.** The authors acknowledge Guoan Zheng for the valuable discussions.